\pdfoutput=1
\documentclass[prl,twocolumn,showpacs,letterpaper,10pt,aps,notitlepage,nofootinbib,superscriptaddress]{revtex4-2}

\usepackage[utf8]{inputenc}
\usepackage{amsfonts}
\usepackage{amsmath}
\usepackage{graphicx}
\usepackage{hyperref}
\usepackage{placeins}
\usepackage{mathrsfs}
\usepackage{bm}
\usepackage[utf8]{inputenc}
\usepackage{multirow}
\usepackage{slashed}
\usepackage{dsfont}
\usepackage{cleveref}
\usepackage[export]{adjustbox}
\hypersetup{colorlinks, linkcolor = [rgb]{0,0.0,0.75}, citecolor = [rgb]{0,0.0,0.75}, urlcolor = [rgb]{0,0.0,0.75}}

\begin{document}

\title{Charm-baryon semileptonic decays and the strange \texorpdfstring{$\bm{\Lambda^*}$}{Lambda} resonances: \texorpdfstring{\\}{} New insights from lattice QCD}

\author{Stefan Meinel}
\affiliation{Department of Physics, University of Arizona, Tucson, AZ 85721, USA}

\author{Gumaro Rendon}
\affiliation{Physics Department, Brookhaven National Laboratory, Upton, NY 11973, USA}

\date{February 22, 2022}

\begin{abstract}
Understanding the properties of the strange $\Lambda^*$ baryon resonances is a long-standing and fascinating problem. $\Lambda_c$ charm-baryon semileptonic weak decays to these resonances are highly sensitive to their internal structure and can be used to test theoretical models. We have performed the first lattice-QCD computation of the form factors governing $\Lambda_c$ semileptonic decays to a $\Lambda^*$ resonance: the $\Lambda^*(1520)$, which has negative parity and spin $3/2$. Here we present the resulting Standard-Model predictions of the $\Lambda_c\to\Lambda^*(1520)\ell^+\nu_\ell$ differential and integrated decay rates as well as angular observables. Furthermore, by combining the recent BESIII measurement of the $\Lambda_c \to X e^+ \nu_e$ inclusive semipositronic branching fraction [\href{https://journals.aps.org/prl/abstract/10.1103/PhysRevLett.121.251801}{Phys.~Rev.~Lett.~{\bf 121}, 251801 (2018)}] with lattice-QCD predictions of the $\Lambda_c \to \Lambda e^+ \nu_e$, $\Lambda_c \to n e^+ \nu_e$, and $\Lambda_c \to \Lambda^*(1520) e^+ \nu_e$ decay rates, we obtain an upper limit on the sum of the branching fractions to all other semipositronic final states. In particular, this upper limit constrains the $\Lambda_c\to\Lambda^*(1405)e^+ \nu_e$ branching fraction to be very small, which may be another hint for a molecular structure of the $\Lambda^*(1405)$.
\end{abstract}

\maketitle

\FloatBarrier

In the quark model, the $\Lambda$ baryons are bound states of one up, one down, and one strange quark, with the light up and down quarks in a flavor-antisymmetric (isospin-0) combination. In addition to the lightest $\Lambda$ baryon with a mass of $1115.683(6)$ MeV and spin-parity $J^P=\frac12^+$, a large number of heavier, short-lived $\Lambda^*$ resonances have been observed in kaon-proton scattering and other processes \cite{Zyla:2020zbs}. Many of these resonances are visible, for example, in the $p K^-$ invariant-mass distribution of  $\Lambda_b \to J/\psi p K^-$ decays analyzed by LHCb -- the same decays that also show pentaquark resonances in the $J/\psi p$ distribution \cite{Aaij:2015tga}. Understanding the properties of the $\Lambda^*$ resonances remains challenging, and little is known quantitatively about their internal structure, even though the underlying fundamental theory -- quantum chromodynamics (QCD) -- is well established. The lowest-mass enhancement with $J^P=\frac12^-$, referred to as the $\Lambda^*(1405)$\footnote{The $^*$ is often omitted when specifying the mass in the name of the particle, but we will continue to use the $^*$ notation to distinguish the QCD-unstable resonances from the QCD-stable ground state.}, likely corresponds to two separate poles in the meson-baryon scattering amplitude \cite{Oller:2000fj,Meissner:2020khl,Mai:2020ltx,Hyodo:2020czb}, leading the Particle Data Group to add a new $\Lambda^*(1380)$ entry, albeit with a two-star rating for ``only fair'' evidence of existence, in its 2020 edition \cite{Zyla:2020zbs}. It has been speculated early on that the $\Lambda^*(1405)$ has a molecular structure \cite{Dalitz:1967fp,Dalitz:1960du}, and there is new evidence for such a structure from an exploratory lattice QCD study of its strange magnetic form factor \cite{Hall:2014uca,Hall:2016kou} (see also Refs.~\cite{Menadue:2011pd,Molina:2015uqp,Gubler:2016viv,Liu:2016wxq,Tsuchida:2017gpb,Pavao:2020zle} for spectroscopic lattice studies).

Quantitative information on the internal structure of the $\Lambda^*$ resonances can also be obtained from the semileptonic charm-baryon decays $\Lambda_c \to \Lambda^* \ell^+ \nu_\ell$  \cite{Ikeno:2015xea}, where the charm quark inside the $\Lambda_c$ decays to a strange quark, leading to the formation of the $\Lambda^*$, together with a positron or positive muon, $\ell^+$, and the associated neutrino, $\nu_\ell$. The decay rates and angular distributions depend on the hadronic structure through the matrix elements \mbox{$\langle \Lambda^*(p^\prime) | \bar{s}\gamma^\mu (1-\gamma_5) c |\Lambda_c(p)\rangle$}; these matrix elements are usually expressed in terms of several independent form factors, which are functions of $q^2=(p-p^\prime)^2$ only. For example, the quark model of Ref.~\cite{Hussain:2017lir}, which treats the $\Lambda^*(1405)$ as a three-quark $uds$ bound state, predicts the $\Lambda_c \to \Lambda^*(1405) \ell^+ \nu_\ell$ decay rate to be about 100 times larger than the study of Ref.~\cite{Ikeno:2015xea}, in which the $\Lambda^*(1405)$ is treated as a dynamically generated meson-baryon molecular state. However, none of the $\Lambda_c \to \Lambda^* \ell^+ \nu_\ell$ decay rates to specific $\Lambda^*$ resonances (other than the ground-state $\Lambda$) have been measured in experiments to date.

Another way to test the accuracy of quark models and other theoretical approaches is to perform first-principles, model-independent calculations directly in the fundamental theory of quantum chromodynamics. At hadronic energy scales, the interactions are too strong to use perturbation theory, but one can perform nonperturbative computations numerically using lattice gauge theory. Lattice-QCD computations have reached sub-percent precision for several quantities \cite{Aoki:2019cca}. Concerning $\Lambda_c$ semileptonic decays, complete lattice-QCD calculations have been published in 2016 for $\Lambda_c \to \Lambda \ell^+ \nu_\ell$ with the lightest $\Lambda$ baryon in the final state \cite{Meinel:2016dqj}, and in 2017 for  $\Lambda_c \to n \ell^+ \nu_\ell$ with the neutron in the final state \cite{Meinel:2017ggx}. So what can be done for the more interesting $\Lambda^*$ final states? Because the $\Lambda^*$'s are unstable under the strong interactions, lattice-QCD calculations are substantially more challenging. To perform the calculations in a completely rigorous way, one would need to actually compute the transition matrix elements to all relevant multi-hadron channels in which the $\Lambda^*$ resonances occur, such as $\Lambda_c \to p K^-  \ell^+ \nu_\ell$ and $\Lambda_c \to \Sigma \pi  \ell^+ \nu_\ell$, and then analytically continue these matrix elements to the locations of the $\Lambda^*$ poles at complex center-of-mass energy. The determination of the multi-hadron transition matrix elements is computationally expensive, and involves additional steps to relate the interacting finite-volume states on the lattice with the asymptotic, noninteracting infinite-volume states. The necessary theoretical formalism has been developed for the case of two-hadron channels \cite{Lellouch:2000pv,Briceno:2014uqa,Briceno:2015csa}, but so far has been applied numerically only to the simpler $\pi\gamma \to \pi\pi$ transition \cite{Briceno:2015dca,Alexandrou:2018jbt}.

There is, however, one $\Lambda^*$ resonance that is sufficiently narrow such that directly identifying the lowest finite-volume state on the lattice with the resonance is expected to be a very good approximation.\footnote{This approach works well for non-$S$-wave resonances and at zero spatial momentum only, in which case the extra energy level caused by the narrow resonance will be far below all scattering states for typical lattice sizes.} This is the $\Lambda^*(1520)$, the lightest resonance with $J^P=\frac32^-$, which has a mass of approximately 1519 MeV and a width of approximately 16 MeV \cite{Zyla:2020zbs}, and is responsible for the sharpest and highest peak in $\Lambda_b \to J/\psi p K^-$ \cite{Aaij:2015tga}. Even this resonance was in fact elusive for lattice calculations in the past \cite{Engel:2012qp}, because local interpolating fields are unsuitable. In Ref.~\cite{Meinel:2020owd}, we have obtained clear signals for the $\Lambda^*(1520)$ in lattice QCD using an interpolating field with a spatial structure corresponding to a total quark orbital angular momentum of $L=1$, and we have used this field to compute the form factors relevant for the rare bottom-baryon decays $\Lambda_b \to \Lambda^*(1520)\ell^+\ell^-$. As described in detail in an accompanying longer article \cite{Meinel:2021mdj}, we have now applied these methods to the determine also the charm-to-strange  $\Lambda_c \to \Lambda^*(1520)$ transition form factors. Unlike for $\Lambda_b \to \Lambda^*(1520)$, the lower mass of the $\Lambda_c$ allowed us to determine the form factors in the entire kinematic range relevant for the semileptonic decays $\Lambda_c \to \Lambda^*(1520) \ell^+ \nu_\ell$: $m_\ell^2 \leq q^2 \leq (m_{\Lambda_c}-m_{\Lambda^*(1520)})^2$. In the following, we present the resulting Standard-Model predictions for these decays. For comparison, we also show predictions for $\Lambda_c \to \Lambda \ell^+ \nu_\ell$ and $\Lambda_c \to n \ell^+ \nu_\ell$, using form factors from the lattice-QCD calculations in Refs.~\cite{Meinel:2016dqj} and \cite{Meinel:2017ggx}.

\begin{figure*}
 \includegraphics[width=0.49\linewidth]{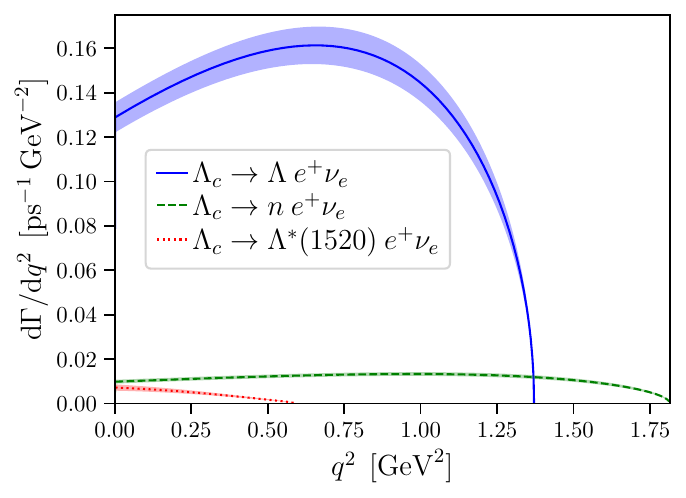} \hfill \includegraphics[width=0.49\linewidth]{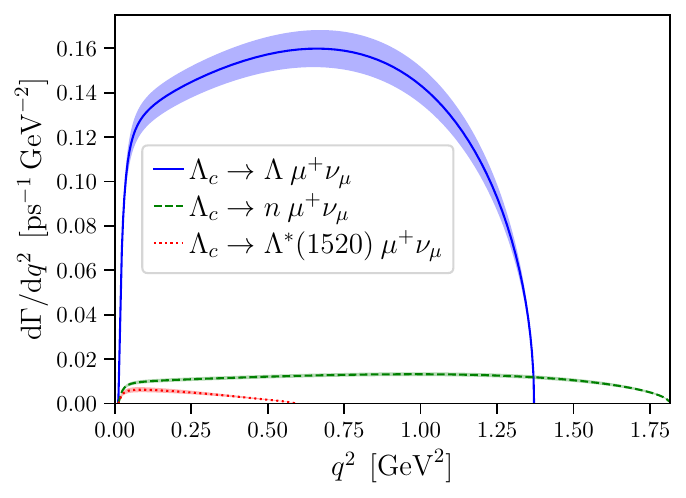}

 \includegraphics[width=0.49\linewidth]{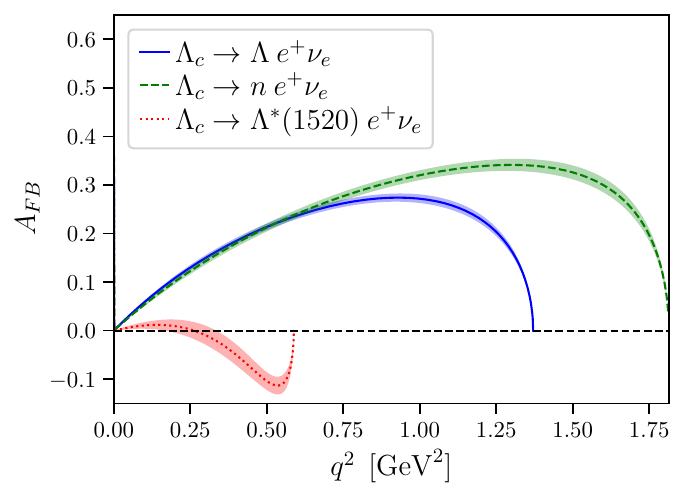} \hfill \includegraphics[width=0.49\linewidth]{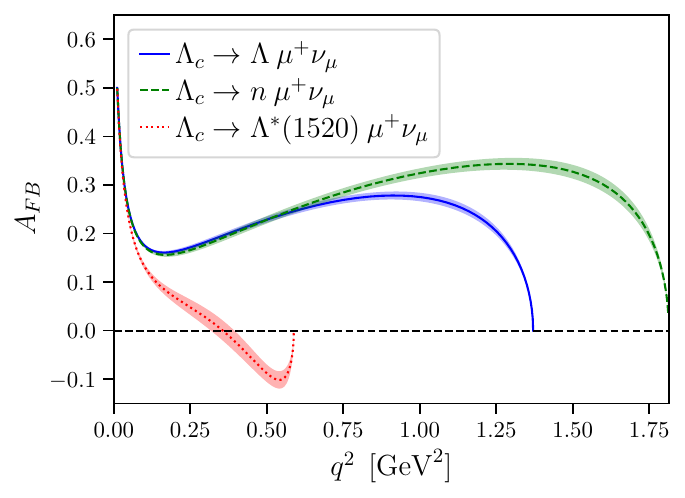}

  \includegraphics[width=0.49\linewidth]{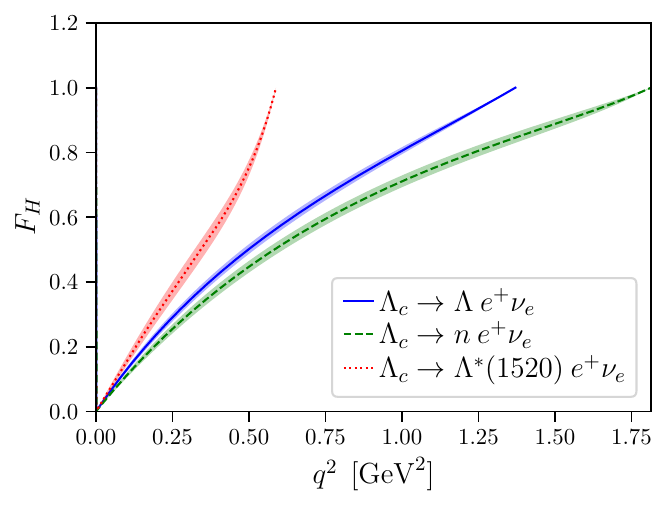} \hfill \includegraphics[width=0.49\linewidth]{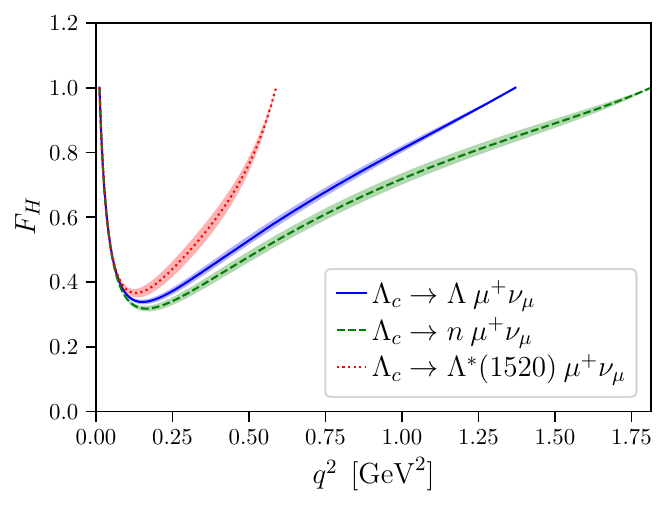}

 \caption{\label{fig:observables}Top: The differential (with respect to $q^2$) decay rates of $\Lambda_c$ baryons to $\Lambda \ell^+ \nu$, $n \ell^+ \nu$, and $\Lambda^*(1520) \ell^+ \nu$. Center and bottom: the two angular observables describing the distribution of the helicity angle of the charged lepton, for unpolarized $\Lambda_c$. All observables were calculated in the standard model of particle physics using lattice QCD. }
\end{figure*}

We consider the two-fold differential decay distributions with respect to the dilepton invariant-mass-squared $q^2=(p-p^\prime)^2$ and $\cos\theta_\ell$, where $\theta_\ell$ is the angle of the $\ell^+$ in the dilepton rest frame with respect to the direction of flight of the dilepton system in the $\Lambda_c$ rest frame. In the Standard Model, these decay distributions can be written as
\begin{equation}
 \frac{\mathrm{d}^2\Gamma^{(F)}}{\mathrm{d}q^2\:\mathrm{d}\cos\theta_\ell} = A^{(F)} + B^{(F)} \cos\theta_\ell + C^{(F)} \cos^2\theta_\ell,
\end{equation}
where the superscript $F$ is used to label the baryon in the final state [$F=\Lambda^*(1520),\Lambda,n$], and the coefficients $A^{(F)}$, $B^{(F)}$, $C^{(F)}$ are functions of $q^2$ only \cite{Gutsche:2015mxa,Boer:2018vpx,Boer:2019zmp}. We take the expressions for $A^{(F)}$, $B^{(F)}$, and $C^{(F)}$ in terms of the form factors from Ref.~\cite{Boer:2018vpx} for the $J^P=\frac32^-$ final state $F=\Lambda^*(1520)$ and from Ref.~\cite{Boer:2019zmp} for the $J^P=\frac12^+$ final states $F=\Lambda,n$, and appropriately modify them for the case of decays to $\ell^+\nu_\ell$ \cite{Korner:1989qb,Kadeer:2005aq,Gutsche:2015mxa} (flipping the sign of the non-$m_\ell^2$-suppressed terms in $B^{(F)}$). The integral over $\cos\theta_\ell$ yields the $q^2$-differential decay rate
\begin{equation}
 \frac{\mathrm{d}\Gamma^{(F)}}{\mathrm{d}q^2} = 2 A^{(F)} + \frac23 C^{(F)},
\end{equation}
and the normalized angular observables we consider are the forward-backward asymmetry
\begin{equation}
 A_{FB}^{(F)} = \frac{B^{(F)}}{\mathrm{d}\Gamma^{(F)}/\mathrm{d}q^2}
\end{equation}
and the ``flat term''
\begin{equation}
 F_{H}^{(F)} = \frac{2 (A^{(F)} + C^{(F)})}{\mathrm{d}\Gamma^{(F)}/\mathrm{d}q^2}.
\end{equation}
The differential decay rates are shown at the top of Fig.~\ref{fig:observables}. Note that the decay to the neutron involves the transition of the charm quark to a down quark and the rate is proportional to $|V_{cd}|^2$,  while the decay rates to the $\Lambda$ and $\Lambda^*$ are proportional to $|V_{cs}|^2$, with the magnitudes of the Cabibbo-Kobayashi-Maskawa (CKM) matrix elements $|V_{cd}|=0.22500(54)$, $|V_{cs}|=0.97344(12)$ \cite{UTfit}. Our lattice-QCD prediction for the integrated $\Lambda_c\to\Lambda^*(1520) \ell^+\nu_\ell$ decay rates divided by $|V_{cs}|^2$ (reported in this form to remove dependence on this external Standard-Model parameter) are
\begin{equation}
 \frac{\Gamma(\Lambda_c\to\Lambda^*(1520) \ell^+\nu_\ell)}{|V_{cs}|^2}
 = \left\{\begin{array}{ll} 0.00267(39)(18)\:{\rm ps}^{-1}, & \ell=e, \\
                            0.00239(34)(16)\:{\rm ps}^{-1}, & \ell=\mu. \end{array}\right. \label{eq:Gamma}
\end{equation}
The $\Lambda_c\to\Lambda^*(1520) e^+\nu_e$ rate is approximately 75 times smaller than the $\Lambda_c\to\Lambda e^+\nu_e$ rate \cite{Meinel:2016dqj}. 
Multiplying $\Gamma$ by the measured $\Lambda_c$ lifetime of $\tau_{\Lambda_c}=0.2024(31)\:{\rm ps}$ \cite{Zyla:2020zbs} yields the branching fractions
\begin{equation}
 \mathcal{B}(\Lambda_c\to\Lambda^*(1520) \ell^+\nu_\ell)
 = \left\{\begin{array}{ll} 0.0512(82)(8)\:\%, & \ell=e, \\
                            0.0458(72)(7)\:\%, & \ell=\mu, \end{array}\right. \label{eq:BF}
\end{equation}
where the first uncertainty is from the lattice calculation and the second uncertainty is from $\tau_{\Lambda_c}$.

The angular observables are shown in the center and lower panels of Fig.~\ref{fig:observables}. It is interesting to see that the forward-backward asymmetries of $\Lambda_c\to\Lambda^*(1520) \ell^+\nu_\ell$ are negative near $q^2_{\rm max}$, in contrast to the $J^P=\frac12^+$ final states. Note that $F_H\to 1$ and $A_{FB}\to 0$ for $q^2\to q^2_{\rm max}$ as a result of exact endpoint relations \cite{Hiller:2021zth}.

When comparing the $\Lambda_c\to\Lambda^*(1520) \ell^+\nu_\ell$ observables to those evaluated using the quark-model form factors of Ref.~\cite{Hussain:2017lir}, we find the differential decay rates from lattice QCD to be about 3 times higher near $q^2_{\rm max}$ and about 1.5 times lower near $q^2=0$, while the integrated rates from lattice QCD are only about 10 percent lower and therefore in agreement within the uncertainties. Our lattice-QCD results for the angular observables $F_H$ are also close to the quark-model prediction. For the forward-backward asymmetry $A_{FB}$, we find the location of the zero-crossings predicted by lattice QCD to be at lower values of $q^2$ (the quark model predicts the zero-crossings to occur at $q^2\approx0.41\:{\rm GeV}^2$ for $\ell=e$ and  $q^2\approx0.44\:{\rm GeV}^2$ for $\ell=\mu$).

Finally, note that the BESIII collaboration has recently measured the inclusive branching fraction $\mathcal{B}(\Lambda_c\to X e^+\nu_e)$, which refers to the sum of $\Lambda_c$ branching fractions to arbitrary hadrons or combinations of hadrons, $X$, together with a positron and neutrino, obtaining \cite{Ablikim:2018woi}
\begin{equation}
 \mathcal{B}(\Lambda_c\to X e^+\nu_e)_{\rm BESIII} = 3.95(34)(9)\:\%. \label{eq:inclusiveBESIII}
\end{equation}
For comparison, the sum of the lattice-QCD predictions for the branching fractions to the $\Lambda$, $n$, $\Lambda^*(1520)$ final states is\footnote{BESIII has also measured $\mathcal{B}(\Lambda_c \to \Lambda e^+ \nu_e)$ \cite{Ablikim:2015prg}. The result is consistent with the lattice-QCD prediction but has a larger uncertainty.}
\begin{eqnarray}
\nonumber &&\mathcal{B}(\Lambda_c\to \Lambda e^+\nu_e)_{\rm LQCD} + \mathcal{B}(\Lambda_c\to n e^+\nu_e)_{\rm LQCD}  \\
\nonumber && +\mathcal{B}(\Lambda_c\to\Lambda^*(1520) e^+\nu_e)_{\rm LQCD}  \\
\nonumber &=& 3.85(20)(6)\:\%  +  0.415(27)(6)\:\%  +  0.0512(82)(8)\:\% \\
 &=&  4.32(23)(7)\:\%, \label{eq:inclusiveLQCD}
\end{eqnarray}
where the first uncertainty is from the lattice calculations (assumed to be fully correlated among the different final states) and the second uncertainty is from $\tau_{\Lambda_c}$.

By subtracting Eq.~(\ref{eq:inclusiveLQCD}) from Eq.~(\ref{eq:inclusiveBESIII}) we obtain the following result for the sum of all other semipositronic branching fractions:
\begin{equation}
 \mathcal{B}(\Lambda_c\to X e^+\nu_e)_{X\neq \Lambda,n,\Lambda^*(1520)} = -0.37 \pm 0.43\,\%.
\end{equation}
Applying the Feldman-Cousins procedure to a Gaussian distribution that is constrained to be non-negative \cite{Feldman:1997qc}, this translates to upper limits of 
\begin{eqnarray}
 \nonumber \mathcal{B}(\Lambda_c\to X e^+\nu_e)_{X\neq \Lambda,n,\Lambda^*(1520)}  < 0.15\,\% \text{ at }68\%\:\:\text{CL}, \\
 \nonumber  \mathcal{B}(\Lambda_c\to X e^+\nu_e)_{X\neq \Lambda,n,\Lambda^*(1520)}  < 0.39\,\% \text{ at }90\%\:\:\text{CL}. \\
\end{eqnarray}
In particular, these results also represent upper limits on the branching fraction to the mysterious $\Lambda^*(1405)$. The quark models of Refs.~\cite{Hussain:2017lir} and \cite{Li:2021qod}, which treat the  $\Lambda^*(1405)$ as a three-quark bound state, predict $\mathcal{B}(\Lambda_c\to\Lambda^*(1405) e^+\nu_e)$ to be $0.24\%$ and $0.31\%$, respectively, both above the 68\%-confidence-level upper limit. In contrast, Ref.~\cite{Ikeno:2015xea}, in which the $\Lambda^*(1405)$ arises as a molecular state in unitarized chiral perturbation theory, predicts $\mathcal{B}(\Lambda_c\to\Lambda^*(1405) e^+\nu_e)$ to be in the range from $0.002\%$ to $0.005\%$. If the preference for a small $\mathcal{B}(\Lambda_c\to X e^+\nu_e)_{X\neq \Lambda,n,\Lambda^*(1520)}$ is strengthened with a higher-precision measurement of the inclusive branching fraction and a higher-precision lattice calculation or measurement of $\mathcal{B}(\Lambda_c\to \Lambda e^+\nu_e)$ [whose uncertainty dominates the uncertainty of Eq.~(\ref{eq:inclusiveLQCD})], this will provide another hint for an exotic structure of the $\Lambda^*(1405)$.

\FloatBarrier

\begin{acknowledgments}
\textit{Acknowledgments:} We thank M.~Bordone, S.~Descotes-Genon, G.~Hiller, C.~Marin-Benito, J.~Toelstede, D.~van Dyk, and R.~Zwicky for discussions. We are grateful to the RBC and UKQCD Collaborations for making their gauge field ensembles available. SM is supported by the U.S. Department of Energy, Office of Science, Office of High Energy Physics under Award Number D{E-S}{C0}009913. GR is supported by the U.S. Department of Energy, Office of Science, Office of Nuclear Physics, under Contract No.~D{E-S}C0012704 (BNL). The computations for this work were carried out on facilities at the National Energy Research Scientific Computing Center, a DOE Office of Science User Facility supported by the Office of Science of the U.S. Department of Energy under Contract No. DE-AC02-05CH1123, and on facilities of the Extreme Science and Engineering Discovery Environment (XSEDE) \cite{XSEDE}, which is supported by National Science Foundation grant number ACI-1548562. We acknowledge the use of Chroma \cite{Edwards:2004sx,Chroma}, QPhiX \cite{JOO2015139,QPhiX}, QLUA \cite{QLUA}, MDWF \cite{MDWF}, and related USQCD software \cite{USQCD}.
\end{acknowledgments}

\providecommand{\href}[2]{#2}\begingroup\raggedright\endgroup

\end{document}